\documentclass[a4paper,11pt]{article} 

\usepackage{amsmath, amsthm, amssymb, marvosym}

\usepackage{enumerate}
\usepackage{hyperref}

\usepackage{color}

\usepackage{graphicx}
\usepackage{epstopdf}
\usepackage{pgf}

\usepackage{soul}

\usepackage[normalem]{ulem}

\usepackage{subcaption}

\newcommand{\bbC}{{\mathbb C}}

\newcommand{\bbR}{{\mathbb R}}

\newcommand{\ep}[1]{\mathrm{e}^{#1}}

\newcommand{\ket}[1]{| #1 \rangle}
\newcommand{\bra}[1]{\langle #1 |}
\newcommand{\tr}{\operatorname{tr}}

\newcommand{\id}{\mathbf{1}}

\renewcommand{\d}{\mathrm{d}}

 \newtheorem{thm}{Theorem}

 \newtheorem{lemma}[thm]{Lemma}
  
  \newtheorem{rem}{Remark}

\newcommand{\comment}[1]{}

\author{Ian Chi,  Martin Fraas, Tina Tan \\
\small{UC Davis} }

\begin{document}

\title{Poincar\'{e} disk as a model of squeezed states of a harmonic oscillator}
\date{\today}%
\maketitle

\begin{abstract}
Single-mode squeezed states exhibit a direct correspondence with points on the Poincar\'e disk. In this study, we delve into this correspondence and describe the motions of the disk generated by a quadratic Hamiltonian. This provides a geometric representation of squeezed states and their evolution. We discuss applications in bang-bang and adiabatic control problems involving squeezed states.
\end{abstract} 

\section{Introduction}

Geometric representations of quantum states constitute essential tools of quantum mechanics. Widely employed examples include the Bloch sphere representation for qubits \cite{bloch, feynman} and the complex plane representation for coherent states \cite{glauber}, both extensively detailed in quantum mechanics textbooks. The Bloch sphere captures the states of a two-level system in a one-to-one manner. For systems beyond two levels, a comprehensive representation in low-dimensional space becomes impossible. However, it is still possible to represent a sub-manifold, such as the above mentioned coherent states. Another option is to represent classes of states, e.g. states of two qubits modulo local unitary \cite{avron}. These geometric representations serve as frameworks, enabling us to conceptualize quantum states and their dynamic evolution through classical analogies. A comprehensive reference for geometric representations is a book \cite{bengtsson} and a review \cite{avronreview}.

This study delves into the representation of ground states of quadratic Hamiltonians, commonly referred to as squeezed states. We show that squeezed states can be represented on the Poincar\'e disk model of hyperbolic geometry. Within this mapping, the evolution of squeezed states corresponds to motions of the disk. We demonstrate the utility of the representation by solving some associated control problems.

The representation of squeezed states on the Poincar\'e was first explored in \cite{bachmann}, where it was used to visualize adiabatic crossing of a point where infinitely many eigenvalues collide. In this paper we describe this correspondence in more details. For the sake of completeness we reproduce some results from \cite{bachmann} in the appendix.

\section{Setup and results}

We consider the family of quadratic Hamiltonians parameterized by $\omega \geq 0 $ and $\alpha \in \mathbb{C}$ given by
\begin{equation}
\label{eq:Hamiltonian}
H = \omega a^* a + \frac{\alpha}{2} a^2 + \frac{\bar{\alpha}}{2} {a^*}^2.
\end{equation}
The operators $a, a^*$ are irreducible representations of the canonical commutation relation
$
    [a,a^*]=1.
$
Our results will be independent of a choice of representation. A {\it squeezed state} $\psi$ is a state satisfying 
\begin{equation}\label{z}
( a+ z a^*)\psi=0
\end{equation}
for some $z \in\bbC$. The
state $\psi$ exists if and only if $|z|<1$, in which case it is
determined up to a phase by $z$. In the Schr\"{o}dinger representation, solutions of the equation
are multiples of
\begin{equation*}
\psi(x)=\ep{-\frac{1+z}{1-z}\frac{x^2}{2}}.
\end{equation*}
The solutions are indeed  in $L^2(\bbR)$ if and only if the condition $|z| <1$ holds, see \cite{bachmann}.

We will use $[\psi(z)]$ to denote the complex line spanned by solutions of Eq.~(\ref{z}) and by $\psi(z)$ or $\ket{\psi(z)}$ a vector in $[\psi(z)]$ of unit norm. 
The manifold
\begin{equation*}
    \mathcal{C} := \{[\psi(z)], |z| < 1\}
\end{equation*}
of the complex projective space of $\mathcal{H}$ is formed by the set of all squeezed states. 
The mapping 
\begin{equation*}
\begin{split}
    F:D&\to \mathcal{C}\\
    z&\mapsto[\psi(z)]
\end{split}
\end{equation*}
associating a point in the open unit disk $D \subset \mathbb{C}$ with the squeezed state $[\psi(z)]$ is one-to-one.

The disk $D$ equipped with distance 
\begin{equation*}
    d(z,w)=\ln \frac{|\bar{z} w -1|+|z-w|}{|\bar{z} w -1|-|z - w|}
\end{equation*}
is known as the Poincar\'e disk model of hyperbolic geometry. 
In particular, the hyperbolic distance from a point $z$ to the origin is
\begin{equation}
\label{d_h}
    d(0,z)=\ln \frac{1+|z|}{1-|z|}.
\end{equation} 
The Poincar\'e metric, the metric induced by $d_h$, is
\begin{equation*}
    g_h = 4\frac{1}{(1 -|z|^2)^2} |\d z|^2.
\end{equation*}

In Section~\ref{sec:mobius} we will discuss the Poincar\'e disk model in more detail. In particular, we will introduce motions of the Poincar\'{e} disk and their invariants: hyperbolic circles, horocycles and hypercycles.

Our main focus is on the pullback of dynamics generated by $H$ on the Poincar\'e disk. The qualitative properties of the dynamics depend on the spectral properties of the Hamiltonian in Eq.~(\ref{eq:Hamiltonian}). There are three cases that we express in terms of parameters $\omega, \alpha$. 
\begin{enumerate}[i)]
\item The \emph{stable case:} If $\omega > |\alpha|$, then the Hamiltonian has the pure point spectrum $\sigma = \lambda \left( n + \frac{1}{2}
\right) - \frac{\omega}{2} \quad (n=0,1,2,\ldots)$.
\item The \emph{free case:} If $\omega = |\alpha|$, then the Hamiltonian has the absolutely continuous spectrum $\sigma = [-\omega, \infty)$.
\item The \emph{unstable case:} If $\omega < |\alpha|$, then the spectrum is $(-\infty, \infty)$.
\end{enumerate}

\begin{figure}
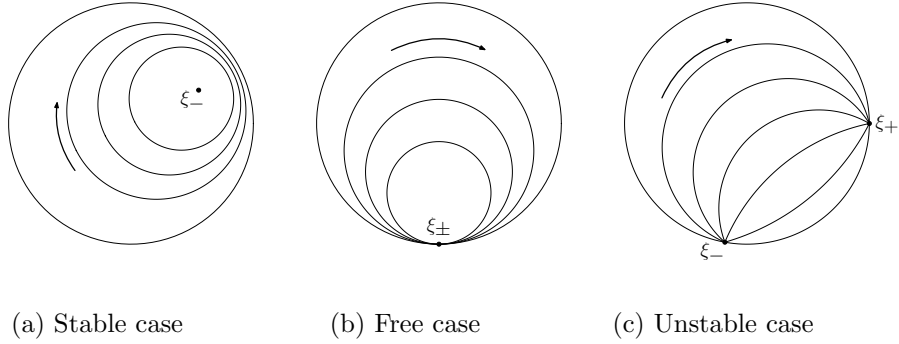
%
\begin{subfigure}[t]{0.3\textwidth}
    \scalebox{0.3}{\input{intro_stable.pgf}}
    \caption{Stable case}
    \label{fig:hyperbolic circles intro}
\end{subfigure}%
~
\begin{subfigure}[t]{0.3\textwidth}
    \scalebox{0.3}{\input{intro_free.pgf}}
    \caption{Free case}
    \label{fig:horocycles intro}
\end{subfigure}%
~
\begin{subfigure}[t]{0.3\textwidth}
    \scalebox{0.3}{\input{intro_unstable.pgf}}
    \caption{Unstable case}
    \label{fig:hypercycles intro}
\end{subfigure}
    \caption{Trajectories of $z(t)$.}
    \label{fig:invariant_curves}
\end{figure}

Let $\lambda = \sqrt{\omega^2 - |\alpha|^2}$. The complex points 
\begin{equation} \label{eq:fixed_points}
    \xi_\pm=\frac{\omega\pm\lambda}{\alpha}
\end{equation}
will play an important part in describing the dynamics of the evolution $U(t) = \ep{-i t H}$ generated by the Hamiltonian in Eq.~(\ref{eq:Hamiltonian}). In the stable case, $\xi_- \in D$ represents the ground state of $H$ and $\xi_+ \notin D$. In the free and unstable cases, we have $|\xi_\pm| =1$, where the points are degenerate in the free case. We will show in Section~\ref{sec:evolution} that the evolution $U(t)$ leaves the manifold of squeezed states $\mathcal{C}$ invariant.  The pullback of $U(t)$ gives a motion $M(t)$ of the Poincar\'{e} disk. Explicitly,
\begin{equation}
\label{eq:M(t)}
M(t)z = F^{-1} \left( [U(t) \psi(z)] \right).
\end{equation}
We write $z(t) = M(t) z$. By the above, $z(t)$ satisfies $[U(t) \psi(z)] = [\psi(z(t))]$. We now describe the geometry of the motion.
\begin{thm}\label{theorem:circles}
Let $C = \{z(t) | t \in [-\infty,\infty) \}$. Then 
\begin{enumerate}[i)]
\item\label{case:z(t)1} The stable case: $C$ is a hyperbolic circle with center $\xi_-$ (Fig.~\ref{fig:hyperbolic circles intro}). 
\item\label{case:z(t)2} The free case: $C$ is a horocycle tangent to the unit circle at the point $\xi_- = \xi_+$ (Fig.~\ref{fig:horocycles intro}).
\item\label{case:z(t)3} The unstable case: $C$ is a hypercycle intersecting the unit circle at the points $\xi_\pm$ (Fig.~\ref{fig:hypercycles intro}).   
\end{enumerate}
In cases~\ref{case:z(t)1}) and \ref{case:z(t)2}), the motion is clockwise, and in case~\ref{case:z(t)3}) the trajectory travels from $\xi_-$ to $\xi_+$. 
\end{thm}
The proof of this theorem and more details about the motion are in Section~\ref{sec:evolution}.

In Section~\ref{sec:control}, we will discuss several control problems. In particular, we derive the set of states reachable by a sequence of pulses obtained by switching $n$ times between two quadratic Hamiltonians $H_0, H_1$.

Finally, in Appendix~\ref{geometry} we show that the manifold of squeezed states $\mathcal{C}$ equipped with the Fubini-Study metric is a realization of hyperbolic geometry. 

\section{M\"{o}bius Transformations}
\label{sec:mobius}

We provide an overview of M\"{o}bius transformations, see, for example, \cite{needham}. A M\"{o}bius transformation is a mapping  \(f:\mathbb{C} \rightarrow \mathbb C\) of the form
\[f(z) = \frac{az+b}{cz+d}\]
for \(a,b,c,d\in\mathbb C\). Such transformations form a group of all conformal mappings of \(\mathbb C\cup \{ \infty \} \) to itself. 
We write the matrix $[f]$ associated with $f$ as 
\begin{equation*}
[f]=\begin{bmatrix} a & b \\ c & d \end{bmatrix}.
\end{equation*}
The mapping $f \to [f]$ is a group homomorphism.

A non-identity M\"{o}bius transformation has two fixed points, counted with multiplicity, given by 
\begin{equation}\label{eq:fixedPoint}
    \xi_{\pm}=\frac{(a-d)\pm\sqrt{(a-d)^2+4bc}}{2c}.
\end{equation}
We use the same letter as in Eq.~(\ref{eq:fixed_points}) since the points $\xi_\pm$ in Eq.~(\ref{eq:fixed_points}) will be fixed points of a specific M\"{o}bius transformation.

In this paper, we only consider M\"{o}bius transformations that preserve the open unit disk $D$. This is the case if and only if the transformation takes the form
\begin{equation}
    M(z)=\frac{az+b}{\bar{b} z+\bar{a}}
\end{equation}
for some $a,b \in \mathbb{C}$ with $|a| > |b|$. There are three classes of such transformations: elliptic, parabolic, and hyperbolic. The positions of the fixed points $\xi_\pm$ determine the class. An automorphism of $D$ is 
\begin{enumerate}[i)]
    \item elliptic if $\xi_- \neq \xi_+$ and only one fixed point is in $D$, in which case the points are related by circle inversion, i.e. $\xi_+ = 1/\overline{\xi_-}$,
    \item parabolic if there is a single fixed point $\xi_- = \xi_+$ on the unit circle, and
    \item hyperbolic if $\xi_- \neq \xi_+$ and both fixed points are on the unit circle.
\end{enumerate}
The invariant curves of elliptic, parabolic, and hyperbolic automorphisms are hyperbolic circles, horocycles, and hypercycles respectively. These curves are Euclidean circles or circular arcs in $D$. In the Poincar\'e disk model, a Euclidean circle corresponds to
\begin{enumerate}[i)]
    \item a hyperbolic circle if it is contained within $D$,
    \item a horocycle if it is tangent to the unit circle,
    \item a hypercycle if it intersects the unit circle at two points at non-right angles, and
    \item a hyperbolic line if it intersects the unit circle orthogonally.
\end{enumerate}

Let $M$ be any M\"{o}bius transformation. We will use a decomposition
\begin{equation*}
    M = S \circ T \circ S^{-1}
\end{equation*}
with $S$ given by
\begin{equation}\label{eq:S_nondegen}
    S = \frac{z-\xi_-}{z-\xi_+}
\end{equation}
in the elliptic and hyperbolic cases, and
\begin{equation}\label{eq:S_degen}
    S = \frac{1}{z-\xi_\pm}
\end{equation}
in the parabolic case. The transformation $T$ is a rotation $z \mapsto e^{i\theta}z$ in the elliptic case, a translation $ z \mapsto z + \beta$ in the parabolic case, and a dilation $z \mapsto r z$ in the hyperbolic case for some non-zero $\theta, r \in \mathbb{R}$ and $\beta \in \mathbb{C}$. The compositions $S \circ M  = T \circ S$ give the normal form of $M$. 

\section{Motions of squeezed states}
\label{sec:evolution}
In this section, we prove Theorem~\ref{theorem:circles}. We start by showing that $U(t) = e^{-itH}$ indeed preserves the unit disk and give an explicit formula for the corresponding M\"{o}bius transformation $M(t)$ in Eq.~(\ref{eq:M(t)}).

The evolution $U(t)$ generated by the Hamiltonian (\ref{eq:Hamiltonian}) preserves the linear space spanned by $a, a^*$. Let $a(t) = U^*(t) a U(t)$ be the Heisenberg evolution of $a$. Taking the time derivative and using $
    [H,a] = -\omega a-\bar\alpha a^*$, we get a system of equations

\begin{equation} \label{eq:system}
    \begin{bmatrix}
        \dot a(t) \\
        \dot a^*(t)
    \end{bmatrix}
    = i
    \begin{bmatrix}
        -\omega & -\bar\alpha\\
        \alpha & \omega
    \end{bmatrix}
    \begin{bmatrix}
        a(t) \\
        a^*(t)
    \end{bmatrix}.
\end{equation}

The solution with $a(0) = a$ and $a^*(0) = a^*$ is $a(t) = u(t) a + v(t) a^*$, where $u(t), v(t)$ are complex functions whose form will be discussed later.
This implies that for any complex function $z(t)$,
$$
U^*(t) (a + z(t)a^*) U(t) \psi(z) = \left( (u(t) + z(t) \overline{v(t)}) a + (v(t) + z(t) \overline {u(t)}) a^* \right) \psi(z).
$$
We now pick the function $z(t)$ so that the combination of $a, a^*$ on the right hand side annihilates $\psi(z)$, i.e.
$$
\frac{v(t) + z(t) \overline {u(t)}}{u(t) + z(t) \overline{v(t)}} = z,
$$
and conclude that $(a + z(t)a^*) U(t) \psi(z) = 0$. Explicitly, we showed that 
$$
[U(t) \psi(z)] = \psi(z(t)),
$$
with 
$$
z(t) = \frac{z u(t) - v(t)}{-z \overline{v(t)} + \overline{u(t)}}.
$$
Since $U(t)$ is unitary, $\psi(z(t))$ is in $L^2(\bbR)$, so $|z(t)| <1$ and the M\"{o}bius transformation $M(t) : z \to z(t)$ preserves the Poincar\'{e} disk. We will also check this explicitly after computing $u(t), v(t)$.

The Hamiltonian evolution $U(t)$ forms a group and hence the pullback $M(t) = F^{-1} U(t)$ is also a group, where
$$
M(s) \circ M(t) = M(s+t), \quad s,t \in \mathbb{R}.
$$
It follows that the fixed points of the M\"{o}bius transformations $M(s)$ are independent of $s$. 

We now discuss the mapping $M(t)$ in each of the three cases.

\subsection{The stable case}
\label{sec:stable}
Let $\lambda = \sqrt{\omega^2 - |\alpha|^2}$. The solution of the system of equations~(\ref{eq:system}) is 
\begin{equation*}
    \begin{bmatrix}
        a(t) \\
        a^*(t)
    \end{bmatrix}
    = \frac{a(\omega-\lambda)+\bar\alpha a^*}{2\lambda} e^{i \lambda t}
    \begin{bmatrix}
        -1\\
        \frac{\omega+\lambda}{\bar\alpha}
    \end{bmatrix}
    +\frac{a(\omega+\lambda)+\bar\alpha a^*}{2\lambda} e^{-i \lambda t}
    \begin{bmatrix}
        1\\
        \frac{\lambda - \omega}{\bar\alpha}
    \end{bmatrix}.
\end{equation*}
This implies that
$$
    z(t) = \frac{\bar\alpha(e^{i\lambda t}-e^{-i\lambda t}) + z((\lambda-\omega)e^{i\lambda t}+(\omega+\lambda)e^{-i\lambda t})}{-\alpha z(e^{i\lambda t}-e^{-i\lambda t}) + ((\lambda-\omega)e^{-i\lambda t}+(\omega+\lambda)e^{i\lambda t})}.
$$
The fixed points are then, c.f. Eq.~(\ref{eq:fixedPoint}),
$$
    \xi_\pm=\frac{\omega\pm\lambda}{\alpha}.
$$
Since there are two distinct fixed points $\xi_\pm$ which are circle inversions of one another, the transformation $M(t)$ is elliptic. We have the decomposition 
\begin{equation}\label{eq:z_compos}
    z(t) = (S \circ T \circ S^{-1})z
\end{equation}
where $S$ is given by Eq. (\ref{eq:S_nondegen}) and $T$ is the rotation $z \mapsto e^{-2i\lambda t}z$. This implies that $C = \{z(t) | t \in [-\infty,\infty) \}$ is a hyperbolic circle centered at $\xi_-$, with $\psi(\xi_-)$ being the ground state of $H$. The trajectory of $z(t)$ makes a full revolution in time $\pi /  \lambda$.

\subsection{The free case}

The solution of the system of equations~(\ref{eq:system}) gives
$$ z(t) = \frac{-i \bar{\alpha}t - (1 - i \omega t)z}{i \alpha t z - (1+i \omega t)}$$
which has a single fixed point
\[ \xi = \frac{\omega}{\alpha}.\]
This is a parabolic transformation and its decomposition (\ref{eq:z_compos}) can be written by taking $S$ as defined in Eq.~(\ref{eq:S_nondegen}) and $T$ to be the translation $z \mapsto z -i \alpha t$.

This shows that $C = \{z(t) | t \in [-\infty,\infty) \}$ is a horocycle touching the unit circle at $\xi$.

\subsection{The unstable case}
Let $\gamma = \sqrt{|\alpha|^2 - \omega^2}$. The solution of the system of equations~(\ref{eq:system}) is  
\begin{equation*}
z(t) = \frac{-\bar{\alpha}(e^{\gamma t}-e^{- \gamma t}) + z((\omega + i \gamma)e^{ \gamma t}-(\omega-i \gamma)e^{-\gamma t})}{\alpha z(e^{\gamma t}-e^{- \gamma t}) - ((\omega - i \gamma)e^{ \gamma t}-(\omega+i \gamma)e^{-\gamma t})}
\end{equation*}
with fixed points
\begin{equation}
    \xi_\pm =  \frac{\omega \pm i \gamma}{\alpha},
\end{equation}
where $z(t)\rightarrow \xi_\pm$ as $t \rightarrow \pm \infty$. The fixed points lie on the unit circle so this is a hyperbolic transformation whose decomposition (\ref{eq:z_compos}) is given by taking $S$ as defined in Eq. (\ref{eq:S_degen}) and $T$ to be the dilation $ z \mapsto e^{2 \gamma t}z$.

This shows that $C = \{z(t) | t \in [-\infty,\infty) \}$ is a hypercycle connecting $\xi_-$ to $\xi_+$.

\section{Applications to control theory}
\label{sec:control}

The questions we study are motivated by \cite{burgarth} that studied reachable set in a similar control problem. Our methods allows us to give a full geometric characterization of reachable sets in several control problems associated with squeezed states.

\subsection{Bang-bang control}
We consider two quadratic Hamiltonians $H_0, H_1$ and the control sequence given by 
\begin{equation}
\label{eq:bang-bang}
\psi(t)=S_n \cdots S_j \cdots S_2 S_1 \psi(0)
\end{equation}
where each operator $S_j$ acts over a time interval of length $\Delta t_j$. Alternating between the time evolution operators corresponding to $H_0$ and $H_1$, we have 
\begin{equation*}
S_j=
    \begin{cases}
        e^{-iH_0 \Delta t_j} & \text{for } j=2k+1, \\
        e^{-iH_1 \Delta t_j} & \text{for } j=2k,
    \end{cases}
\end{equation*}
and $t = \Delta t_1 + \Delta t_2 + \dots + \Delta t_n$. The operator \(U_k\) is defined as the product
\begin{equation*}
    U_k =  S_k\cdots S_{2} S_{1}.
\end{equation*}

The control problem we consider involves evolving an initial squeezed state $[\psi(z_{0})]$ into the target state $[\psi(z_f)]$ for given $z_{0}$ and $z_f$ using bang-bang control (\ref{eq:bang-bang}). We consider several cases of $H_0, H_1$.

\subsubsection{The stable case}
We consider two stable Hamiltonians, i.e. $\omega > |\alpha|$,
\begin{align*}
H_0&=\omega a^* a, \\
H_1&=\omega a^* a + \frac{\alpha}{2} a^2 + \frac{\bar{\alpha}}{2}a^{*2},
\end{align*}
where, for simplicity, we have chosen $H_0$ to be the standard harmonic oscillator.
We are specifically interested in finding the minimum number of times that we need to use the Hamiltonian $H_1$ in order to move the initial state into the target state with complete freedom to pick $\Delta t_j$. Taking $n = 2k+1$, we want to find minimal $n$ for which there exists a sequence $\Delta t_1, \dots, \Delta t_n$ that solves the control problem. The answer depends only on the modulus of $z_0$ and $z_f$. Let $\xi$ denote the fixed point in $D$ associated with $H_1$. We define sequences $R_{2 k +1}, r_{2k+1}$ recursively as 
\begin{equation}
\label{eq:maxR}
R_{2k+1}=\frac{(1+|\xi|^2) R_{2k-1} + 2|\xi|}{2 |\xi| R_{2k-1} + (1+ |\xi|^2)}, \quad R_1 = |z_0|,
\end{equation}
and
\begin{equation}
\label{eq:minr}
r_{2k+1}=
    \begin{cases}
        \frac{(1+|\xi|^2) r_{2k-1} - 2|\xi|}{-2 |\xi| r_{2k-1} + (1+ |\xi|^2)}, & \text{if } r_{2k-1} > \frac{2|\xi|}{1+|\xi|^2} \\
        0, & \text{otherwise.} 
    \end{cases}, \quad r_1 = |z_0|. 
\end{equation}

\begin{thm}\label{thm:bang-bang}
The control problem $[\psi(z_f)] = [U_{2k +1} \psi(z_0)]$ has a solution if and only if
$$
r_{2k+1} \leq |z_f| \leq R_{2k+1}.
$$ 
\end{thm}
\begin{proof}
Using the mapping of squeezed states to the Poincar\'{e} disk, the problem is equivalent to the study of motions $F^{-1} \circ U_k$ on the disk. By Theorem~\ref{theorem:circles}, the trajectories $F^{-1} \circ S_k$ correspond to circles centered at the origin for odd terms and hyperbolic circles with hyperbolic center $\xi$ for even terms. Thus for each $S_{2k+1}$, there is some maximum and minimum $|z|$-value reachable by that step, which we denote $R_{2k+1}$ and $r_{2k+1}$ respectively, see Fig.~\ref{fig:hcircle_control}.

\begin{figure}
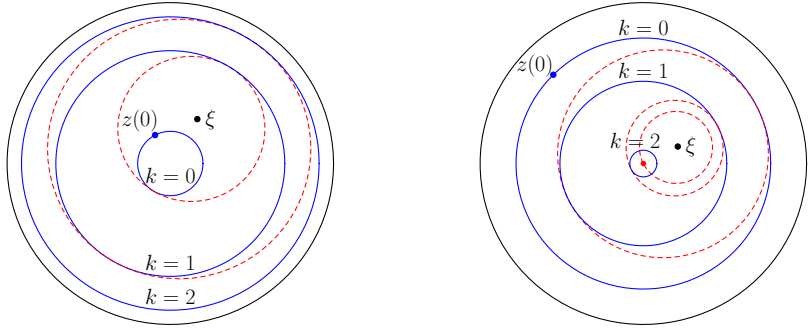

    \scalebox{0.4}{\input{stable_control_max.pgf}} 
    \scalebox{0.4}{\input{stable_control_min.pgf}}
    \caption{Illustrations of how to reach the maximum radius (left) and minimum radius (right).}
    \label{fig:hcircle_control}
\end{figure}
The initial terms are given by $R_1=r_1=|z_0|.$ Subsequent terms are computed as follows:
Given a Euclidean circle of radius $R_{2k-1},$ the next term $R_{2k+1}$ is determined by the point on the circle with maximal hyperbolic distance from $\xi$, which is $p=-R_{2k-1} \xi / |\xi|$. Let $q$ be the point antipodal to $p$ on the hyperbolic circle centered at $\xi$ passing through $p$. Then $R_{2k+1}=|q|$, which can be computed  by converting the hyperbolic distance $d(0, q) = d(0,\xi) + d(p, \xi)$ to Euclidean distance using Eq.~(\ref{d_h}). We obtain the recursion Eq.~(\ref{eq:maxR}).

For the minimal reachable radius, we get Eq. (\ref{eq:minr}) using similar geometric considerations as for the maximal radius. The recursion holds as long as the the trajectory does not intersect with the hyperbolic circle centered at $\xi$ passing through the origin. Once these intersect, the origin is reachable so the following term is zero.

\end{proof}

\begin{rem}

The recurrence equation for the maximum radius, Eq.~(\ref{eq:maxR}), is a hyperbolic Mobius transformation $R_{2k+1} = M(R_{2k-1})$ with an explicit formula $R_{2k+1} = M^{\circ k} R_1$ and a corresponding matrix
\begin{equation*}
[M] = \begin{bmatrix} 1+|\xi|^2 & 2|\xi| \\ 2|\xi| & 1+|\xi|^2
\end{bmatrix}.
\end{equation*}
We can diagonalize $[M]$ with
\begin{equation*}
U=\frac{1}{\sqrt{2}} \begin{bmatrix} 1 & 1 \\ 1 & -1 \end{bmatrix}, 
\quad
D=\begin{bmatrix} (1+|\xi|)^2 & 0 \\ 0 & (1-|\xi|)^2 \end{bmatrix}.
\end{equation*}
Using $[M]=U D U^{-1}$ and $[M^{\circ k}]=U D^k U^{-1}$, we have
$$
R_{2k+1} = \frac{ ( 1 + \Delta^{2k})R_1 + 1 - \Delta^{2k}}{(1 - \Delta^{2k}) R_1 +1 + \Delta^{2k}}, \quad \Delta = \frac{1 - | \xi|}{1 + |\xi|} .
$$
For large $k$, $\Delta^{2k} <<1$ and we get an asymptotic formula
\begin{equation*}
    1- R_{2k+1} \approx  2\left(\frac{1-R_1}{1+R_1}\right) \Delta^{2k}.
\end{equation*}
\end{rem}

\subsubsection{The free case}
Consider the Hamiltonians $H_0, H_1$ given by
\begin{align*}
H_i&= \omega_i a^* a + \frac{\alpha_i}{2} a^2 + \frac{\bar{\alpha}_i}{2}a^{*2}
\end{align*}
where $\omega_i = |\alpha_i|$ and the fixed points are $\xi_i= \omega_i/\alpha_i$. In order to describe the set of reachable states, we use arc-polygons, i.e. polygons whose edges are Euclidean circular arcs. 

\begin{figure}
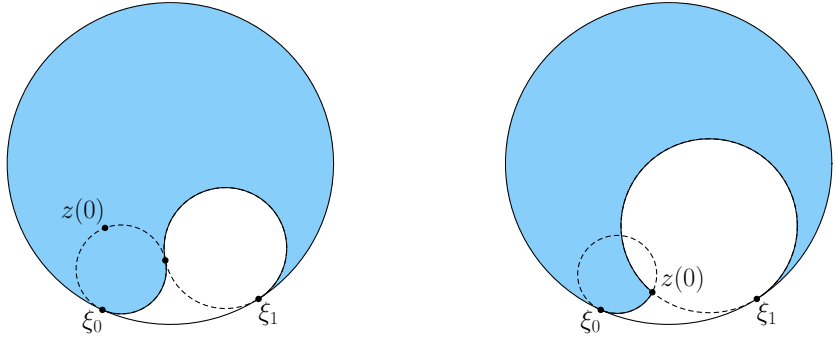

    \scalebox{0.4}{\input{free_control_1.pgf}}
 \hfill
    \scalebox{0.4}{\input{free_control_2.pgf}}
    \caption{Set of reachable states (blue) in the free case after applying $H_0, H_1$. 
    }
    \label{fig:horocycle_control}
\end{figure}

Evolving some initial point $z_0$ under $H_0$ results in a trajectory along a horocycle through $\xi_0$. Applying $H_1$ then extends the reachable set to an arc-triangle (Fig. \ref{fig:horocycle_control}). In all cases, we have that
\begin{itemize}
    \item $\xi_0$ is a vertex with internal angle $\pi$,
    \item $\xi_1$ is a vertex  with internal angle zero, and
    \item the arc of the unit circle traversed clockwise from $\xi_0$ to $\xi_1$ is an edge.
\end{itemize}
The remaining vertex and edges are determined by the point where the horocycle through $z(0)$ and $\xi_0$ intersects with the horocycle through $\xi_1$ tangent to it. If the intersection point can be reached by evolving $z(0)$ under $H_0$, then the intersection point is the third vertex. Otherwise, the third vertex is $z(0)$. The remaining two edges are given by horocycle arcs connecting the third vertex to $\xi_0$ and $\xi_1$. The entire open disk is reachable with the sequence $H_0, H_1, H_0$.

\subsubsection{The unstable case}

\begin{figure}
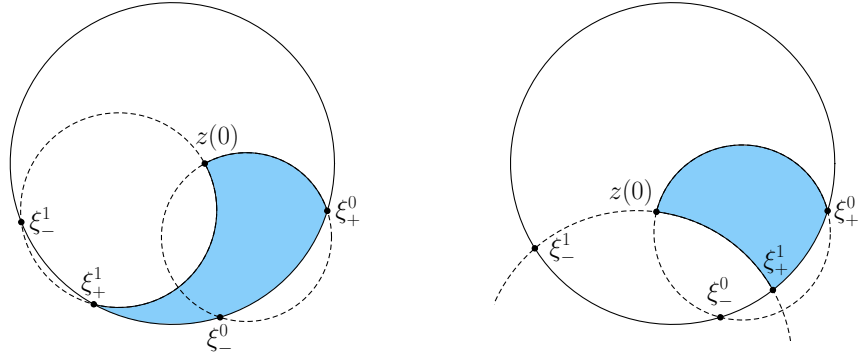
%
\begin{subfigure}[t]{0.5\textwidth}
    \scalebox{0.4}{\input{unstable_control_4.pgf}}
    \caption{Non-overlapping pairs of fixed points}
    \label{fig:unstable_non_overlap}
\end{subfigure}%
~
\begin{subfigure}[t]{0.3\textwidth}
    \scalebox{0.4}{\input{unstable_control_3.pgf}}
    \caption{Overlapping pairs of fixed points}
    \label{fig:unstable_overlap}
\end{subfigure}%
~
\caption{Set of reachable states (blue) in the unstable case after applying $H_0, H_1$.}
    \label{fig:hypercycle_control_1}
\end{figure}

For Hamiltonians $H_0, H_1$ where $\omega_i < |\alpha_i|$, the entire disk is reachable only when pairs of fixed points do not overlap (Fig.  \ref{fig:unstable_non_overlap}), in which case any point in $D$ can be reached with the sequence $H_0, H_1, H_0$. Otherwise, the final set is given by some arc-polygon with vertices $z(0)$, $\xi_+^0$, and $\xi_+^1$ (Fig. \ref{fig:unstable_overlap}).

The selection of $H_0, H_1$ that we present here does not encompass all cases, but we hope it convinces the reader that the geometric map gives a versatile method to think about control problems with squeezed states.

\subsection{Adiabatic control}
Provided $\omega^2 > |\alpha|^2$, the stable case, the Hamiltonian $H \equiv H(\omega, \alpha)$ in Eq.(\ref{eq:Hamiltonian}) has a discrete spectrum with ground state $[\psi(\xi_-)]$. Recall that
$$
\xi_\pm = \frac{\omega \pm \sqrt{\omega^2 - |\alpha|^2}}{\alpha}
$$
are the fixed points of the motion of the Poincar\'{e} disk generated by the Hamiltonian. 

We consider the family of Hamiltonians $H(t)$ corresponding to parameters $\omega(t), \alpha(t)$ that depend smoothly on $t \in [0,1]$. Let $U_\epsilon(t)$ be the unitary generated by the slowly driven Schr\"{o}dinger equation
$$
\epsilon \dot{U}(t) = H(t) U(t), \quad U(0) = \id.
$$
We assume that the parameters satisfy the constraint $\omega(t)^2 > |\alpha(t)|^2$. The ground state $[\psi(\xi_-(t))]$ of $H(t)$ is then protected by a gap, and a basic result of adiabatic theory \cite{Born, Kato} gives the limit
$$
\lim_{\epsilon \to 0} [U_\epsilon(t) \psi(\xi_-(0))] =  [\psi(\xi_-(t))].
$$
For the motion $M_\epsilon(t)  = F^{-1} \circ U_\epsilon(t)$ on the Poincar\'{e} disk, this implies that 
$$
\lim_{\epsilon \to 0} M_\epsilon(t) \xi_-(0) = \xi_-(t).
$$

\begin{figure}[h]
    \centering
    \includegraphics[scale=0.25]{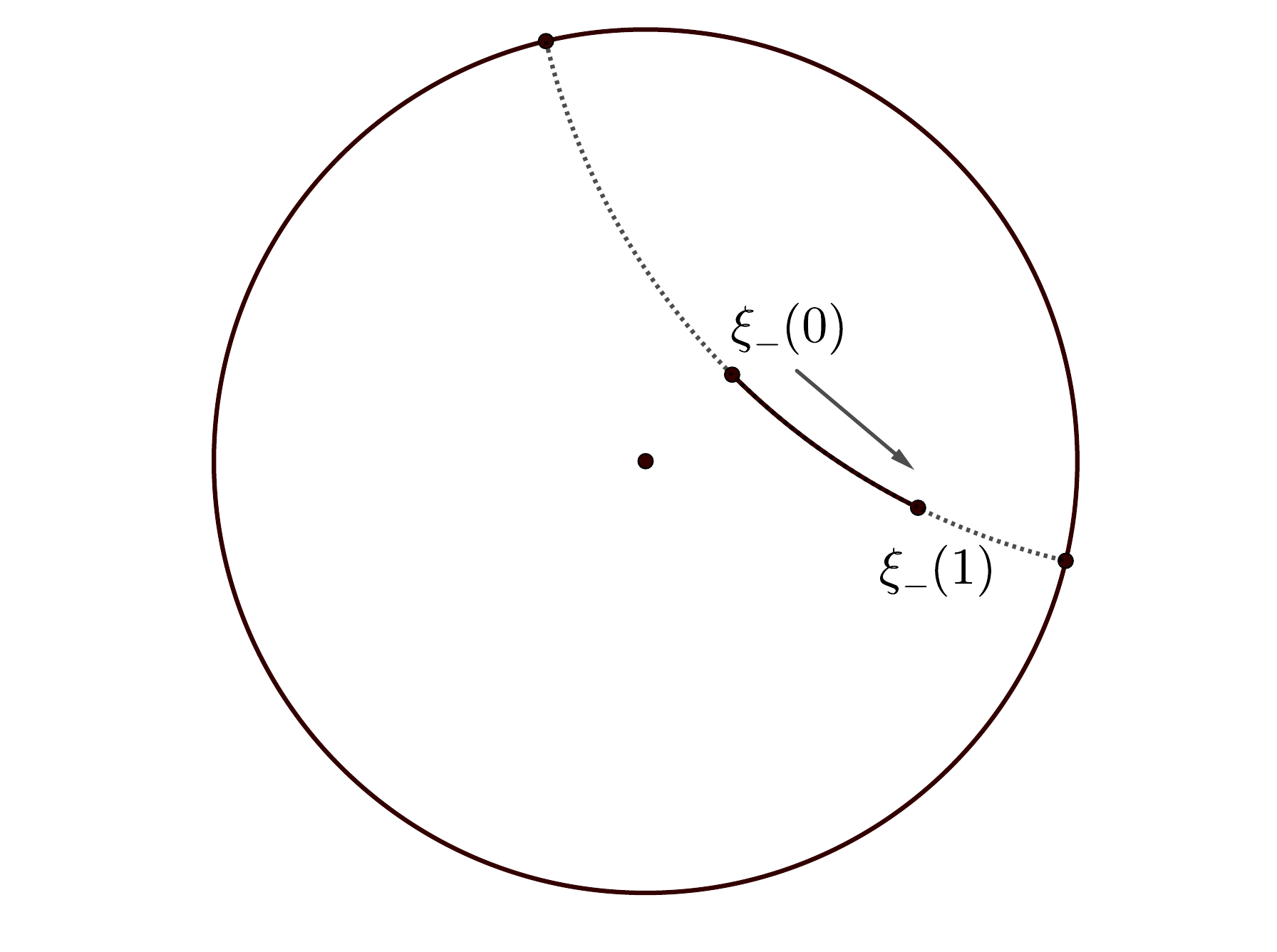}
    \caption{ For linear interpolation, the fixed point $\xi_-$ traces a hyperbolic line.}  
    \label{fig:adiabatic-control-pic}
\end{figure}

The idea of adiabatic control theory is to move the initial state $[\psi(\xi_-(0))]$ into the target state $[\psi(\xi_-(t))]$ by slowly changing the corresponding Hamiltonian $H(t)$. In the limit of infinitely slow driving, the motion traces the curve $\xi_-(t)$ on the disk. Our goal is to describe this curve geometrically (Fig.~\ref{fig:adiabatic-control-pic}).


 We restrict our attention to a particular example. We fix $\omega$ and pick $\alpha(t) = (1-t) \alpha_0 + t \alpha_1$, a linear interpolation between an initial value $\alpha_0$ and a final value $\alpha_1$. 
 \begin{thm}
For the linear interpolation described above, $\xi_-(t)$ traces a hyperbolic line segment. 
 \end{thm}

\begin{proof}
Recalling that $\xi_\pm(t)$ are circular inverses of each other, it suffices to show that $\xi_\pm(t)$ lies on a circle, and orthogonality with the unit circle follows. A rotation $\alpha \mapsto e^{is} \alpha$ rotates the corresponding fixed points as $\xi_{\pm}(t) \mapsto e^{-is} \xi_{\pm}(t)$. Since rotations preserve circles, we can without loss of generality assume that $\alpha(t)$ is a line parallel with the real axis, i.e. has the form
$$
\alpha(t) = (1-t) a_0 + t a_1 + i b.
$$ 
Computing $| \xi_\pm(t) + i\omega/b|^2$, we get
$$
\left| \frac{\omega \pm \sqrt{\omega^2 - |\alpha(t)|^2}}{\alpha(t)}  + \frac{i \omega}{b} \right|^2 = \frac{\omega^2}{b^2} -1.
$$
The right hand side is positive by the assumption that $\omega^2 > |\alpha|^2$, so $\xi_\pm(t)$ indeed lies on a circle with center $-i \omega/b$ and radius $\sqrt{(\omega^2/b^2)-1}$.
\end{proof}

\appendix
\section{Geometry of squeezed states}
\label{geometry}

We will use an alternative formula for the hyperbolic distance, 
\begin{equation}
\label{d_h-a}
d_h(z,w) = \mathrm{arcosh}(1 + 2 \delta(z,w)), \quad \delta(z,w)=\frac{ |z-w|^2}{(1-|z|^2)(1- |w|^2)}.
\end{equation}
We recall the Poincar\'e metric, the metric induced by $d_h$, 
$$
g_h = 4\frac{1}{(1 -|z|^2)^2} |\d z|^2.
$$

The projective space of $\mathcal{H}$ has a natural Fubini-Study metric induced by the Hilbert-Schmidt distance 
$$
\| P - Q\|^2 = 4 \tr((P - Q)^2)
$$
between rank one projections $P$ and $Q$. The factor of $4$ is a convenient normalization. The sub-manifold of squeezed states $\mathcal{C}$ inherits the Fubini-Study metric 
$$
g_{FS} = 4 \tr( d P(z) d P(z)),
$$
where $P(z) = \ket{\psi(z)} \bra{\psi(z)}$ are projections on $[\psi(z)]$. Again, the factor of $4$ is non-standard normalization that fits our problem.

These two manifolds are congruent.
\begin{thm}\label{thm:geometry}
The map $F$ is an isometric immersion of $(D, g_h)$ into $(\mathcal{C}, g_{FS})$. In particular, the manifold of squeezed states is a model of hyperbolic geometry.
\end{thm}

This theorem is not new, see e.g. \cite{bachmann}. For completeness we include the proof. We follow \cite{bachmann} with only small changes.

We compute the Hilbert-Schmidt distance of squeezed states. 
\begin{lemma}
 For any $z,w \in D$, we have 
 $$
 \tr{\left ((P(z) - P(w))^2\right)} = 2(1 - \left(1 + \delta(z,w)\right)^{-\frac{1}{2}}).
 $$
\end{lemma}
\begin{proof}
In the proof we will use the standard Fock basis $\ket{n}$ obtained from $\ket{0}:= \psi(0)$, and by the recursive relation $a^* \ket{n} = \sqrt{n+1} \ket{n+1}$.
    We claim that 
    \begin{equation}
    \label{eq:4.1}
    \psi(z) = (1 - |z|^2)^{\frac{1}{4}} \tilde{\psi}(z), \quad \tilde{\psi}(z) = e^{-\frac{1}{2} z {a^*}^2} \psi(0).
    \end{equation}
    Indeed, using 
    $$
    e^{-\frac{1}{2} z {a^*}^2} a e^{\frac{1}{2} z {a^*}^2} = a + z a^*,
    $$
    we check that
    $$
    (a + z a^*) \tilde{\psi}(z) =0.
    $$
    To compute the normalization, we expand $\tilde{\psi}(z)$ in a Taylor series,
    $$
    \tilde{\psi}(z) = \sum_{n=0}^\infty \left( -\frac{1}{2} \right)^n \frac{z^n}{n!} \sqrt{(2 n)!} \ket{2n}.
    $$
    Then 
    $$
    \langle \tilde{\psi}(z) | \tilde{\psi}(w) \rangle = \sum_{n=0}^\infty \frac{1}{4^n} \frac{(\bar{z} w)^n}{n!^2} (2n)! = (1 - \bar{z}w)^{-\frac{1}{2}},
    $$
    and using this for $z=w$ shows (\ref{eq:4.1}). To compute the Hilbert-Schmidt distance we use 
    $$
 \tr\left( {(P(z) - P(w))^2} \right)= 2 - 2 \tr(P(z) P(w)),
 $$
 and
 $$
 \tr(P(z) P(w)) = |\langle \psi(z) | \psi(w) \rangle|^2= (1-|z|^2)^{\frac{1}{2}} (1-|w|^2)^{\frac{1}{2}} |1- \bar{z}w|^{-1}.
 $$
 The expression in the lemma then follows from the identity
 $$
 |1- \bar{z}w|^2 = |z-w|^2 + (1- |z|^2)(1-|w|^2),
 $$
 and the definition of $\delta(z,w)$.
\end{proof}
Taylor expansion implies that
$$
\tr\left((P(z) - P(w))^2\right) = \delta(z,w) + o(\delta(z,w)).
$$
Comparing this with the hyperbolic distance, we get 
$$
\tr\left( ((P(z) - P(w))^2 \right)= \frac{1}{4} d_h(z,w)^2 + o(\delta(z,w)).
$$
This implies that $g_h = g_{FS}$ as claimed in the theorem.

\section*{Acknowledgements} 
M.F. thanks Daniel Burgarth and Carl Chalk for discussions.

\bibliography{references}
\bibliographystyle{abbrv}
\end{document}